\documentclass[final,1p,times]{elsarticle}

\usepackage{amssymb}

\journal{Nuclear Physics A}

\begin{document}

\begin{frontmatter}



\title{A Theoretical Determination of
$N_{nn}/N_{np}$ in Hypernuclear Non--Mesonic Weak Decay}


\author{E. Bauer$^{1}$ and G. Garbarino$^2$}

\address{$^1$Departamento de F\'{\i}sica, Universidad Nacional de La Plata and
IFLP, CONICET\\
C. C. 67, 1900 La Plata, Argentina}

\address{$^2$Dipartimento di Fisica Teorica, Universit\`a di Torino\\
and INFN sezione di Torino, I-10125 Torino, Italy}

\begin{abstract}
The ratio $N_{nn}/N_{np}$ between the number of neutron--neutron
and neutron--proton pairs emitted
in the non--mesonic weak decay of $\Lambda$--hypernuclei is
calculated within a nuclear matter formalism extended to
$^{12}_{\Lambda}$C via the local density
approximation. The single--nucleon emission spectra, $N_p$ and $N_n$, are
also evaluated. Our formalism takes care of both
ground state correlations (gsc) and final state interactions (FSI).
The evaluation of $N_{nn}/N_{np}$
---~which, unlike $\Gamma_n/\Gamma_p\equiv
\Gamma(\Lambda n\to nn)/\Gamma(\Lambda p\to np)$, is an
actual observable quantity in non--mesonic decay~--- is performed
within a fully microscopic model where a proper
treatment of FSI, gsc and ground state normalization is
considered. All the isospin channels contributing
to one-- and two--nucleon induced decays are included.
Our final result for the coincidence number ratio,
$N_{nn}/N_{np}=0.374$, is in agreement with
the KEK--E508 datum, $(N_{nn}/N_{np})^{\rm exp}=0.40\pm 0.10$.
\end{abstract}

\begin{keyword}
$\Lambda$--Hypernuclei \sep Non--Mesonic Weak Decay
\sep Two--Nucleon Induced Decay \sep FSI

\PACS 21.80.+a, 25.80.Pw

\end{keyword}

\end{frontmatter}

\section{Introduction}
\label{intro}
Being one of the main sources of information on
strangeness--changing baryon interactions, the
non--mesonic weak decay of $\Lambda$--hypernuclei has attracted a great
deal of attention for several years. In this decay,
the final, observable product is the
emission of two or more nucleons from the hypernucleus.
The non--mesonic decay width, $\Gamma_{\rm NM}$,
is built up from the one-- and two--nucleon induced
decays, $\Gamma_{1} \equiv \Gamma(\Lambda N \to nN)$ and
$\Gamma_{2} \equiv \Gamma(\Lambda NN \to nNN)$, respectively:
$\Gamma_{\rm NM}=\Gamma_{1}+\Gamma_{2}$.
All possible isospin components are: $\Gamma_{n} = \Gamma(\Lambda n \to nn)$,
$\Gamma_{p} = \Gamma(\Lambda p \to np)$, $\Gamma_{nn} = \Gamma(\Lambda nn \to nnn)$,
$\Gamma_{np} = \Gamma(\Lambda np \to nnp)$ and
$\Gamma_{pp} = \Gamma(\Lambda pp \to npp)$.

One should note that the only observables
in hypernuclear weak decay are the
lifetime $\tau$, the mesonic rates $\Gamma_{\pi^-}=\Gamma(\Lambda \to \pi^- p)$
and $\Gamma_{\pi^0}=\Gamma(\Lambda \to \pi^0 n)$
 and the spectra of the emitted particles (nucleons,
pions and photons). None of the non--mesonic partial decay rates
($\Gamma_n$, $\Gamma_p$, $\Gamma_{nn}$, etc) is an observable from a
quantum--mechanical point of view.
Each one of the possible elementary non--mesonic decays occurs in
the nuclear environment, thus subsequent final state interactions (FSI)
modify the quantum numbers of the weak decay nucleons and
new, secondary nucleons are emitted as well: this prevents
the measurement of any of the non--mesonic partial decay rates.

Among the weak decay observables one has the spectra of
emitted neutrons ($N_{n}$) and protons ($N_{p}$) as well as the
$nn$ and $np$ nucleon coincidence spectra, $N_{nn}$ and $N_{np}$.
In the present contribution we
discuss a microscopic model to take care of the
$N_{nn}/N_{np}$ ratio and of the single--nucleon spectra
for the emission of protons and neutrons. We report results for
$^{12}_{\Lambda}$C. The paper is organized as follows:
in Section~\ref{form} we outline the formalism, results for
$^{12}_{\Lambda}$C are discussed in Section~\ref{results} and
finally, in Section~\ref{conclu}, some conclusions are given.

\section{Formalism}
\label{form}
Using the formalism developed in~\cite{ba07}, the number of
particles, $N_{N}$, and of pair of particles, $N_{NN'}$,
where $N \, (N') = n$ or $p$, emitted in non--mesonic decay
can be written as follows:
\begin{eqnarray}
\label{nn1f}
N_{n} & = & 2 \bar{\Gamma}_{n} +  \bar{\Gamma}_{p} +
3 \bar{\Gamma}_{nn} + 2 \bar{\Gamma}_{np} + \bar{\Gamma}_{pp} +
\sum_{i, \, i'; \, j} N_{j \, (n)} \,
\bar{\Gamma}_{i, i' \rightarrow j}~,\\
\label{np1f}
N_{p} & = & \bar{\Gamma}_{p} + \bar{\Gamma}_{np} + 2
\bar{\Gamma}_{pp} +
\sum_{i, \, i'; \, j} N_{j \, (p)} \, \bar{\Gamma}_{i, i' \rightarrow j}~,\\
\label{nnn1f}
N_{nn} & = & \bar{\Gamma}_{n} + 3 \bar{\Gamma}_{nn}
+ \bar{\Gamma}_{np}+
\sum_{i, \, i'; \, j} N_{j \, (nn)} \, \bar{\Gamma}_{i, i' \rightarrow j}~,\\
\label{nnp1f}
N_{np} & = & \bar{\Gamma}_{p} + 2 \bar{\Gamma}_{np}
+ 2 \bar{\Gamma}_{pp} +
\sum_{i, \, i'; \, j} N_{j \, (np)} \, \bar{\Gamma}_{i, i' \rightarrow j}~,\\
\label{npp1f}
N_{pp} & = & \bar{\Gamma}_{pp} +
\sum_{i, \, i'; \, j} N_{j \, (pp)} \, \bar{\Gamma}_{i, i' \rightarrow j}~.
\end{eqnarray}
All terms which contain the functions $\bar{\Gamma}_{i, i' \rightarrow j}$
represent the action of FSI. We refer the reader to~\cite{ba07} for
a complete explanation of the meaning of these functions.
The factors $N_{j \, (N)}$ are the numbers of nucleons of the type $N$
contained in the multinucleon state $j$.
In the same way, the $N_{j \, (NN')}$'s are the numbers of
$NN'$ nucleon pairs in the state $j$. The
summation over $i, \, i'$ and $j$ runs over all possible initial ($i$ and $i'$)
and final ($j$) states. We have used a normalization for which
$\bar{\Gamma} \equiv \Gamma/\Gamma_{\rm NM}$.

The five quantities $N_{N}$ and $N_{NN'}$ of
Eqs.~(\ref{nn1f})--(\ref{npp1f}) are observables. Five
decay widths, $\Gamma_{N}$ and $\Gamma_{NN'}$, also enter
into these equations. The fact
that FSI cannot be neglected makes it impossible to
invert the above relations in order to obtain expressions for the
decay widths in terms of the observables $N_{N}$
and $N_{NN'}$. By starting from the experimental values of $N_{N}$
and $N_{NN'}$ it is possible to extract the so--called
experimental values for the decay widths, once
a model for the FSI is implemented. This is
done in the hope that several models for FSI will
lead to the same values for the extracted decay widths.

\begin{figure}[h]
\centerline{\includegraphics[scale=0.53]{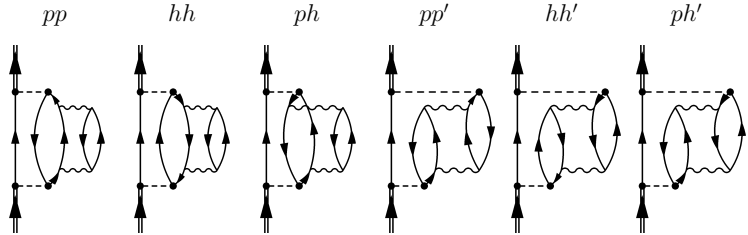}}
\caption{The set of Feynman diagrams considered in this work and
which contain the action of the strong interaction.
The dashed and wavy lines stand for the weak and strong
potentials, $V^{\Lambda N \to NN}$ and $V^{NN}$, respectively.}
\label{fig1}
\end{figure}

In this contribution we adopt a different working method.
Instead of extracting experimental values for the decay widths, we
work out a microscopic model to report results for
$N_{N}$ and $N_{NN'}$ given by Eqs.~(\ref{nn1f})--(\ref{npp1f}).
FSI are modeled by the set of
Feynman diagrams of Fig.~\ref{fig1}. It should
be noted that by considering all the possible time--orderings of these
diagrams, some of the corresponding Goldstone diagrams are pure
gsc contributions (to $\Gamma_{2}$), others are pure
FSI terms and others are interferences
between gsc and FSI contributions.
Pauli exchange terms have not been evaluated.

\section{Results}
\label{results}
Here we report results on the coincidence number ratio, $N_{nn}/N_{np}$,
and on the single--proton and single--neutron emission spectra,
$N_p$ and $N_n$,
as a function of the nucleon kinetic energy.
The calculation is performed in nuclear matter; by the local density
approximation we obtain results for $^{12}_{\Lambda}$C.
The weak transition potential, $V^{\Lambda N \to NN}$, is represented
by the exchange of $\pi$, $\eta$, $K$, $\rho$, $\omega$
and $K^*$ mesons, with the coupling constants and cut--off
parameters deduced from the Nijmegen soft--core interaction NSC97f~\cite{st99}.
For the residual strong interaction,
$V^{N N}$, we use a $\pi+\rho$ potential with the
addition of a $g'$ Landau--Migdal parameter, with $g'=0.7$.

In Table~\ref{table1} we present our results for the
$N_{nn}/N_{np}$ ratio.
The inclusion of FSI brings to a good agreement with data
(without FSI, one would have $N_{nn}/N_{np}=\Gamma_n/\Gamma_p=0.321$).
Conversely, the effect of the quantum interference terms
(which correspond to the diagrams $ph$, $pp'$, $hh'$ and $ph'$ in Fig.~\ref{fig1})
is not important in the determination of $N_{nn}/N_{np}$.
However, from Table~\ref{table2} we see that the individual values of the
interference terms are sizable in comparison with the
dominant ones ($pp$) in both $N_{nn}$ and $N_{np}$.
From Table~\ref{table2} we also see that the
reduction due to the energy and angular cuts suffered by the FSI terms
is much bigger than the reduction for the FSI--free, 1N--ind term. This is
expected, as FSI give a dominant contribution in the low--energy region.
\begin{table}[h]
\begin{center}
\caption{The $N_{nn}/N_{np}$ ratio for $^{12}_{\Lambda}$C. In the
column $(N_{nn}/N_{np})^{0}$ we give results without FSI,
while in the column $(N_{nn}/N_{np})^{\rm no-int}$
values without quantum interference terms are reported.
The KEK datum, from~\cite{ki06}, corresponds to a nucleon kinetic energy
detection threshold $T^{\rm th}_N$ of $30$ MeV and a $NN$ opening angle
region $\cos(\theta_{NN})\leqslant -0.8$.}
\label{table1}
\resizebox*{\textwidth}{!}{
\begin{tabular}{cccccc}   \hline \hline
$~~~~T^{\rm th}_N\, ({\rm MeV})~~~~$ & $\cos(\theta_{NN})$
& $~~\Gamma_{n}/\Gamma_{p}~~$ & $(N_{nn}/N_{np})^{0}$ & $(N_{nn}/N_{np})^{\rm no-int}$
 & $~~N_{nn}/N_{np}~~$ \\  \hline
 $0.$ & $\leqslant 1.$ & $~0.321$  & $~0.321$  & $~0.392$ & $~0.372$    \\
 $30.$ & $\leqslant -0.8$ & $~$  & $~0.332$  & $~0.376$  & $~0.374$  \\ \hline
KEK--E508 & &&  & & $0.40\pm 0.10$    \\ \hline \hline
\end{tabular}}
\end{center}
\end{table}
\begin{table}[h]
\vspace{-1.cm}
\begin{center}
\caption{Partial contributions to $N_{nn}$ and $N_{np}$
for $^{12}_{\Lambda}$C. Predictions for the one--nucleon induced channels
(i.e., without FSI) are given in the column 1N--ind.
For FSI contributions we follow the notation of Fig.~\ref{fig1}.}
\label{table2}
\resizebox*{\textwidth}{!}{
\begin{tabular}{cccccccccc}   \hline \hline
& $T^{\rm th}_N\, ({\rm MeV})$ & $\cos(\theta_{NN})$ & 1N--ind
& $pp$  & $ph$  & $hh$
&$pp'$ & $ph'$  & $hh'$
\\  \hline
$N_{nn}$ & $0.$ & $\leqslant 1.$ & $~0.175$  & $~0.312$  & $-0.049$
 & $~0.303$ & $~0.136$  & $-0.117$ & $~0.137$   \\
& $30.$ & $\leqslant -0.8 $      & $~0.173$  & $~0.060$  & $-0.007$
 & $~0.061$ & $~0.019$  & $-0.012$ & $~0.023$   \\ \hline
$N_{np}$ & $0.$ & $\leqslant 1.$    & $~0.546$  & $~0.903$  & $-0.048$
 & $~0.588$ & $~0.319$  & $-0.141$ & $~0.265$  \\
         & $30.$ & $\leqslant -0.8$ & $~0.520$  & $~0.137$  & $-0.006$
         & $~0.121$ & $~0.054$  & $-0.031$ & $~0.052$   \\
\hline \hline
\end{tabular}}
\end{center}
\end{table}

\begin{figure}[t]
\centerline{\includegraphics[scale=0.43]{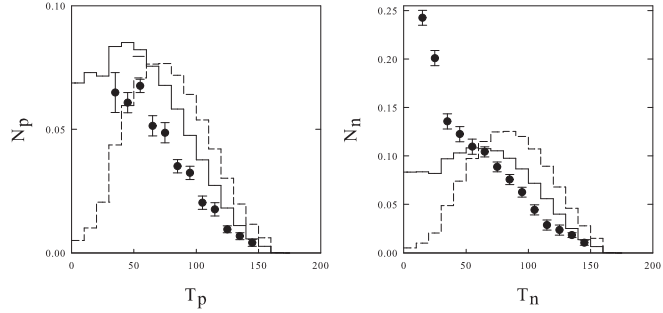}}
\caption{Single--proton ($N_{p}$) and single--neutron ($N_{n}$)
kinetic energy spectra for $^{12}_{\Lambda}C$. With continuous (dashed)
lines we give our results with (without) FSI.
Data are from KEK--E508~\cite{ok04}.}
\label{fig2}
\end{figure}

The single--nucleon spectra obtained with the microscopic model
are shown in Fig.~\ref{fig2}. It is clear that an improvement
has been obtained with respect to the FSI--free predictions.
We observe that it is the action of
the residual strong interaction (through FSI and gsc)
the mechanism which could lead, once additional
many--body terms are considered,
to an agreement with the experimental spectra.

\section{Conclusions}
\label{conclu}
From the experimental spectra of nucleons produced in hypernuclear
non--mesonic weak decay it is not straightforward
to determine the decay widths $\Gamma_{n}$, $\Gamma_{p}$,
$\Gamma_{nn}$, etc:
a theoretical model for FSI is required. In this
contribution, a quantum--mechanical microscopic model for
nucleon emission has been discussed.
It has been shown that with this approach it is possible to obtain a rather
good agreement with experiment for the $N_{nn}/N_{np}$ ratio.
However, we note that one should also reproduce the
individual spectra $N_{nn}$, $N_{np}$, $N_{n}$, $N_{p}$, etc.

From a comparison with data, the predictions of our microscopic model for FSI
shows a clear improvement over the FSI--free results.
Our results suggest that the interference terms are not
important. The present approach requires further improvements:
$i)$ a more realistic residual strong interaction,
$ii)$ the inclusion of Pauli exchange terms for the FSI diagrams and
$iii)$ the study of the effect of the $\Delta(1232)$.
Finally, we note that the microscopic model allows us to set constraints
not only on $V^{\Lambda N \to NN}$ but also on $V^{NN}$.

\section*{Acknowledgments}
One of us (EB) would like to thank the Hyp-X Organizing Committee for
their support to attend the conference.

\end{document}